\documentstyle[12pt]{article}

\def\pd{\partial}
\def\a{\alpha}
\def\b{\beta}
\def\dl{\delta}
\def\s{\sigma}

\def\lam{\lambda}

\def\bg{{\bar g}}
\def\hg{{\hat g}}

\def\hnabla{{\hat \nabla}}

\def\bR{{\bar R}}

\def\bF{{\bar F}}
\def\bG{{\bar G}}

\def\bA{{\bar A}}
\def\bDelta{{\bar \Delta}}
\def\bBox{\stackrel{-}{\Box}}

\def\gm{\gamma}

\def\om{\omega}

\def\sq{\sqrt}
\def\e{\hbox{\large \it e}}

\def\fr{\frac}

\def\arr{\rightarrow}

\def\bb{\begin{equation}}
\def\ee{\end{equation}}
\def\bba{\begin{eqnarray}}
\def\eea{\end{eqnarray}}

\begin{document}

\begin{titlepage}

\begin{tabbing}
   qqqqqqqqqqqqqqqqqqqqqqqqqqqqqqqqqqqqqqqqqqqqqq 
   \= qqqqqqqqqqqqq  \kill 
         \>  {\sc KEK-TH-764}    \\
         \>  {\sc May, 2001} 
\end{tabbing}
\vspace{5mm}

\begin{center}
{\Large {\bf Note on Quantum Diffeomorphism Invariance, 
Physical States and Unitarity}}~\footnote{
Based on the talk given at KEK Workshop 2001, March 12-14.
}
\end{center}

\vspace{5mm}

\centering{\sc Ken-ji Hamada\footnote{E-mail address : 
hamada@post.kek.jp} }

\vspace{3mm}

\begin{center}
{\it Institute of Particle and Nuclear Studies, \break 
High Energy Accelerator Research Organization (KEK),} \\ 
{\it Tsukuba, Ibaraki 305-0801, Japan}
\end{center} 

\vspace{5mm}

\begin{abstract} 
Recently, using a local action satisfying the Wess-Zumino 
condition as a kinetic term of the conformal mode, 
we formulated a four-dimensional quantum geometry 
(4DQG)~\cite{hs,h99,h00}.  
The conformal mode can be treated exactly, 
and it was shown that the part of the effective action related to this 
mode is given by the scale-invariant non-local Riegert action. 
As for the traceless mode, we introduce dimensionless coupling,   
which is a unique gravitational coupling of this theory  
satisfying the conditions of renormalizability and asymptotic freedom.   
Although this theory is asymptotically free, the physical states 
are non-trivial, which should be described as composite fields, 
like the spectrum of 2DQG. 
The possibility that the physical state conditions 
representing background-metric independence conceal ghosts 
is pointed out.
The usual graviton state would be realized when the physical 
state condition breaks down dynamically.       
\end{abstract}
\end{titlepage}  

Einstein's general theory of relativity has been extremely well 
tested,  and it has been believed that the graviton exists, 
at least, classically. However, the existence of a graviton at 
the quantum level has not yet been guaranteed. 
To begin with, it is doubtful whether the usual graviton picture 
can be preserved over the scale of the Planck mass, 
where the usual particle picture seems to gravitationally collapse. 
One of the ideas to overcome such a problem is superstring theory,   
though it is formulated in ten dimensions, so that we must show 
the dynamics of compactification. 
Another possibility may be, as discussed below, a four-derivative 
theory of gravity formulated in a background-metric independent 
manner~\cite{hs,h99,h00,am,amm}. 

Originally, four-derivative theories of gravity were investigated 
to resolve the problem of renormalizability and to avoid a problem 
which arises from the fact that the Einstein-Hilbert action is unbounded 
below~\cite{ud,t77,ft82}. 
However, many unsolved problems have remained in the old 
four-derivative theory. 
One is that the coupling of the conformal mode, $R^2$, with the right 
sign introduced to resolve the unboundedness problem becomes 
asymptotically non-free~\cite{ft82}.\footnote{ 
Note that, in ref.\cite{ft82}, the overall sign is changed to make the 
theory asymptotically free. 
} 
Unitarity becomes unclear~\cite{t77}. 
Furthermore, after 2DQG was solved exactly~\cite{kpz,ddk}, 
it has been noticed~\cite{hs,h99,h00} 
that quantum diffeomorphism invariance, itself, is a problem. 

Since we consider gravity theory over the Planck scale, 
an asymptotically non-free theory is not allowed. 
Thus, we cannot use the $R^2$ term as  
a kinetic term of the conformal mode. 
On the other hand, we know that, as used in 2DQG, the Wess-Zumino (WZ) 
action~\cite{wz}, 
which is a local action obtained by integrating~\cite{p,r,ft84,ds} 
conformal anomalies~\cite{cd,ddi,du},  
can be used as a kinetic term of 
the conformal mode~\cite{ddk,h93,am,amm,hs,h99,h00}. 
Furthermore, this theory can be naturally understood from the viewpoint 
of a quantum diffeomorphism invariance, because the WZ action 
can be interpretated as an action induced from invariant measures 
in order to preserve the diffeomorphism invariance when they are 
replaced with non-invariant measures defined on the background 
metric.  

In a quantum theory of gravity, 
diffeomorphism invariance is promoted to more stringent conditions, 
namely background-metric independence.  
A general coordinate transformation corresponds to an infinitesimal 
change of the background-metric. Background-metric independence 
can be read as the physical state condition, usualy called 
the Hamiltonian-momentum constraints,  
\bb
     \fr{\dl Z}{\dl \hg^{\mu\nu}}
     =\langle {\hat T}_{\mu\nu} \rangle =0, 
\ee
where $Z$ is the partition function and $\hg_{\mu\nu}$ is 
the background metric. This condition can be exactly solved 
in two dimensions~\cite{kpz,ddk,bmi}. 
In this note we point out that the spectrum 
of 4DQG is analogous to that of 2DQG~\cite{bmi}.

\begin{flushleft}
{\bf The model of 4DQG}
\end{flushleft}

The tree action of 4DQG is obtained by combining the WZ action 
and invariant actions as~\cite{hs,h99,h00} 
\bba
   &&{\cal I}= \frac{1}{(4\pi)^2} \int d^4 x \sq{\bg}
        \biggl\{ ~\frac{1}{t^2}{\bar F} + a {\bar F} \phi 
          + 2 b \phi  \bDelta_4 \phi 
          + b \Bigl( {\bar G}-\fr{2}{3} \bBox {\bar R} \Bigr) \phi 
              \nonumber   \\
   && \qquad\qquad\quad
        + \frac{1}{36}(2b+3c){\bar R}^2 \biggr\} 
        + \fr{1}{\hbar}I_{LE}(X,g) ,
                \label{tree}
\eea 
where the metric fields are decomposed as 
$g_{\mu\nu}=\e^{2\phi}\bg_{\mu\nu}$. 
The invariants $F$ and $G$ are the square of the Weyl tensor 
and the Euler density, respectively. 
The operator $\Delta_4$ is the conformally covariant fourth-order 
operator~\cite{r}, which satisfies $\Delta_4 = \e^{-4\phi}\bDelta_4 $ 
locally for a scalar.
The term $I_{LE}$ represents lower derivative actions which  
include actions of conformally invariant matter 
fields, the Einstein-Hilbert action, and the cosmological constant 
term.  Here, note that the Planck constant $\hbar$ appears in front 
of $I_{LE}$, because the metric fields are exactly dimensionless 
so that four-derivative actions of gravity are dimensionless. 
Thus, the four-derivative parts of the tree action are essentially 
quantum effects. 

Since our theory is formulated using the non-invariant measures 
defined on the background metric, 
the conformal mode and the traceless mode must be treated 
as independent fields. 
In the above, we introduce the dimensionless coupling $t$ 
only for the traceless mode as $\bg_{\mu\nu}=(\hg \e^{th})_{\mu\nu}$, 
where $tr(h)=0$~\cite{kkn}, and consider the perturbation of $t$. 
Here, the three coefficients $a$, $b$ and $c$ are not independent 
couplings, where $c$ is associated with a scheme-dependent term.
They can be determined uniquely in the perturbation of $t$ 
by requring diffeomorphism invariance, as briefly reviewed below.
The self-interactions of $\phi$ appear only 
in the lower derivative actions in the exponential form, 
which can be treated exactly, order by order in $t$. 
 
Here, note that we could add extra coupling for the conformal mode, 
$\fr{\zeta}{(4\pi)^2} \sq{g}R^2$, to the tree action ${\cal I}$, 
where $\zeta$ is independent of $a$, $b$ and $c$ as well as $t$ 
and positive for the positivity of the action.\footnote{
This $R^2$ coupling is rather different from that commentted in 
introduction, because here the kinetic term of the conformal mode 
is given by the WZ action, not the part of $R^2$. 
} 
The beta function of this coupling, however, has a positive value,  
$\beta_{\zeta} = \fr{90}{b^2}\zeta^2$~\cite{am}, namely asymptotically 
non-free, so that $\zeta$ must vanish.  
Thus, the asymptotically free dimensionless coupling in the gravity 
sector is only the coupling for the traceless mode, $t$.  
This results in that four-derivative parts of effective action 
related to the conformal mode become scale-invariant. 

One of the important properties of the four-derivative parts of 
the tree action is that it is not diffeomorphism invariant in itself, 
which produces the terms proportional to the forms of conformal 
anomalies~\cite{ddi,du} 
under a general coordinate transformation 
generated by   $\dl g_{\mu\nu} =g_{\mu\lam}\nabla_{\nu}\xi^{\lam} 
+g_{\nu\lam}\nabla_{\mu}\xi^{\lam}$ 
as~\cite{h99,h00} 
\bb
   \dl {\cal I} = \fr{1}{(4\pi)^2} \int d^4 x \sq{\bg} 
      ~\om \Bigl( -a \bF -b \bG - c \bBox \bR \Bigr) , 
              \label{deli}
\ee 
where $\om=-\fr{1}{4}\hnabla_{\lam}\xi^{\lam}$.  

Let us here see how diffeomorphism invariance can be 
realized~\cite{hs,h99,h00}.  
We consider the effective action of this theory, which has the 
following form: 
\bb
   \Gamma = {\cal I}(\phi,\bg) 
              +\sum_A W_A(\bg)  +\cdots,
               \label{eff}
\ee 
where $A$ represents the conformal anomalies $F,~G$ and $\Box R$, 
and the $W_A$'s denote loop corrections associated with them. 
For a while we consider four derivative parts of the effective 
action. In general, the lower-derivative terms receive rather 
complicated corrections.  
Reflecting that the measure defined on the background metric 
changes conformally under a general 
coordinate transformation, $W_A$ transformes 
as follows~\cite{fujikawa}:  
\bb
       \dl W_A(\bg)=\dl_{\om}W_A(\bg)
        =\fr{d_A}{(4\pi)^2}\int \sq{\bg}\om \bA, 
                 \label{anomaly}
\ee 
where $\dl_{\om}\bg_{\mu\nu}=2\om\bg_{\mu\nu}$.
Here, we note that $W_A(\bg)$'s depend only on the traceless mode. 
The actions $W_F(\bg)$ and $W_G(\bg)$ are associated with 
counterterms for $\bF$ and $\bG$, respectively, and  
$W_{\Box R}$ is a scheme-dependent term proportional to $\bR^2$. 
On the other hand, loop diagrams with external $\phi$ fields  
do not diverge~\cite{h99}. 
This is an important feature of our theory. 
Thus, loop corrections with external $\phi$ fields, 
which are now collected in the three-dot symbol in (\ref{eff}), 
become scale-invariant. This corresponds with that there is no 
coupling for the conformal mode. 
In the three-dot symbol, non-anomalous terms, namely conformally 
invariant as well as diffeomorphism invariant terms, are also 
included.  

The coefficients $d_F \equiv f$ and $d_{\Box R}$ in equation 
(\ref{anomaly}) can be determined 
by calculating two-point diagrams of $h$ in the flat background  
using the tree action ${\cal I}$.  
In the momentum space, $W_F+W_{\Box R}$ is given by 
\bba
  && \fr{1}{(4\pi)^2} ~t_r^2 ~h^{\mu}_{~\nu}(p) h^{\lam}_{~\s}(-p) 
     \biggl\{ -\fr{f}{4}
        \biggl( \dl_{\mu\lam}\dl^{\nu\s} p^4 
         -2\dl_{\mu\lam}p^{\nu}p^{\s} p^2 
              \nonumber \\
  && \qquad\qquad\qquad\qquad 
        +\fr{2}{3}p_{\mu}p^{\nu} p_{\lam}p^{\s} \biggr) 
          \log \biggl(\fr{p^2}{\mu^2} \biggr)  
     -\fr{d_{\Box R}}{12}  p_{\mu}p^{\nu} p_{\lam}p^{\s} 
      \biggr\}. 
\eea
The scheme-dependent term can be, for example, fixed as  
$d_{\Box R}=\fr{2}{3}f$ in Duff's scheme~\cite{du}.\footnote{
Here, $u=0$ in refs.\cite{h99,h00} is assumed.
} 
Here, we comment on the scheme-dependent $\bR^2$ term. 
Since the $h^2$ part of the $\bR^2$ term has the same form to 
gauge-fixing term such as $-\chi^{\mu}\pd_{\mu}\pd_{\nu}\chi^{\nu}$ 
in the flat background, where 
$\chi^{\mu}=\pd^{\lam}h^{\mu}_{~\lam}$, this term might be 
gauge-dependent as well as scheme-dependent at the higher loops. 
In any case, the $\bR^2$ terms must cancel out and disappear 
in the effective action from the requirement of diffeomorphism 
invariance.

The coefficient $d_G \equiv e$ can be determined by calculating 
three-point diagrams of $h$ in the flat background.  
The action $W_G(\bg)$ is given by the scale-invariant Riegert 
action~\cite{r} defined on $\bg$, 
\bb     
      \frac{e}{(4\pi)^2}\int d^4 x  
            \sq{\bg} \biggl\{ \fr{1}{8}{\bar {\cal G}} 
             \fr{1}{\bDelta}_4  {\bar {\cal G}} 
               -\frac{1}{18} \bR^2  \biggr\} ,
                \label{wg}
\ee
where ${\cal G} = G - \frac{2}{3}\Box R$.      
Here, the $\bR^2$ term in (\ref{wg}) is necessary to realize 
equation (\ref{anomaly}) and to explain the fact that there is no 
contributions to $e$ from two-point diagrams of $h$. 

The calculable coefficients $f$, $e$ and $d_{\Box R}$ are in general 
given by functions of unknown coefficients $a$, $b$ and $c$.  
The conditions of diffeomorphism invariance represented by 
$\dl \Gamma =0$ can determine these three coefficients uniquely 
as~\cite{hs,h99,h00} 
\bb
     {\tilde a}=f, \quad {\tilde b}=e, \quad c=d_{\Box R} , 
\ee  
so that the anomalies from the tree action (\ref{deli}) and 
the anomalies from loop effects (\ref{anomaly}) cancel out.  
Here, the tildes on $a$ and $b$ stand for the inclusions of 
finite shiftes of these coefficients by loop effects. 
Then, the effective action is obtained by a manifestlly invariant 
form with scheme-independent coefficients $f$ and $e$ as~\cite{h00}  
\bb
    \Gamma = \frac{e}{(4\pi)^2}\int d^4 x  
            \sq{g}  \fr{1}{8}{\cal G} 
             \fr{1}{\Delta}_4  {\cal G}
             +W_F(g)+I_{LE}(g) + \cdots, 
\ee
where the scheme-dependent $W_{\Box R}$ term     
and  the $\bR^2$ term in $W_G(\bg)$ cancel out with the $\bR^2$ 
terms in ${\cal I}$, and $W_F(\bg)$ is replaced with $W_F(g)$. 

The coefficients $f$ and $e$ are scheme-independent. 
They can be expanded by the renormalized 
coupling as $f= \sum_n f_n t_r^{2n}$ and     
$e=\sum_n e_n t_r^{2n}$.   
Here, $f_0$ and $e_0$ have already been computed by one-loop diagrams 
as
\bba
    && f_0 = -\fr{N_X}{120}-\fr{N_W}{40} -\fr{N_A}{10} 
           -\fr{199}{30}+\fr{1}{15} , 
                 \\ 
    && e_0 = \fr{N_X}{360} +\fr{11N_W}{720} +\fr{31N_A}{180} 
           +\fr{87}{20} -\fr{7}{90} , 
              \label{evalue}
\eea 
where the first three contributions of each coefficient come from 
$N_X$ conformal scalar fields, $N_W$ Weyl fermions and $N_A$ gauge 
fields, respectively~\cite{du}. The fourth and the last ones  
come from the traceless mode~\cite{ft82} 
and the conformal mode~\cite{amm}, respectively.   
The coefficients $f_n$ and $e_n$ with $n > 0$ include 
higher-loop corrections. 
Here, $f_0 < 0$ implies that the coupling $t_r$ is asymptotically 
free, namely $\b_t =\fr{f}{2} t_r^3 < 0$.

\begin{flushleft}
{\bf Physical states and unitarity} 
\end{flushleft}

As discussed above, the four-derivative terms of the tree action 
are essentially quantum effects. Furthermore it is not diffeomorphism 
invariant in itself. Thus, in our theory, the terms ``tree'' 
and ``loop'' no longer correspond to the terms ``classical'' 
and ``quantum'', respectively. 
The Born diagrams now imply tree diagrams derived from $L_{LE}$.  
Therefore, we cannot the guess physical asymptotic states 
from the tree action. 
Here, we first review old unitarity arguments~\cite{t77,lw} 
while adding some comments along with our theory. 
Then, we show the possibility that the physical 
state condition conceal ghosts, even at a sufficiently 
high-energy region. 

Diffeomorphism invariance requires that we must take into account 
loop effects when we study physical poles.   
The inverse of propagator of the traceless mode then 
becomes~\footnote{
Here, the Lorentzian siguniture $(-,+,+,+)$ is considered.
} 
\bb
       M^2t_r^2 p^2 +p^4 -\fr{f}{2} t_r^2 p^4 
        \log \biggl( \fr{p^2}{\mu^2} \biggr) , 
\ee
where $M$ is the properly normalized Planck mass.  
The details of the Lorentz indices are omited. 
Hence, the full propagator of the traceless mode has the following form:  
\bb
      \fr{1}{M^2 t_r^2} \fr{1}{p^2} \fr{1}{L(p^2)} , 
      \qquad  L(p^2) = 1-\fr{f}{2M^2} p^2 
            \log \biggl( \fr{p^2}{\mu^{\prime 2}} \biggr) ,
\ee
where $\mu^{\prime}=\mu\e^{-1/f t^2_r}$. 
Here, $1/p^2$ represents physical pole. The condition that 
there is no tachyonic pole is $\mu^{\prime 2} < -\fr{2M^2}{f}\e $. 
The function $L(p^2)$ does not have infra-red catastrophe, 
or $L(0)=1$. Note that the use of the full propagator is now 
due to a symmetrical reason, not due to a kinematical 
reason to treat divergences near the pole. 

Because of the asymptotic freedom, namely $f <0$, $L(p^2)$ does 
not have a real zero. Thus, the ghost pole in the tree action moves 
to a pair of complex pole on the physical sheet~\cite{t77}. 
In relativistic theory, it is known that a vertex decaying 
from real states to such complex-pole states has measure-zero 
contributions~\cite{lw}. 
Thus, there is no vertex decaying to ghosts in diffeomorphism 
invariant theory. It seems that ghosts merely appear as an 
artificial pole to define the perturbation theory using 
non-diffeomorphism invariant vertices. 
The proof of unitarity would be formally given along 
refs.~\cite{lw} using the full propagator in all order of 
the perturbation. 
This feature seems to suggest that background-metric independence 
is imporatnt to unitarity because it may become exact in all orders.
The proof, however, is not rigorous. 
In this note we no longer discuss this idea.   

We here consider what happens at a sufficientlly high-energy 
region, $| p^2| \gg M^2$, because there is a possibility to detect 
ghosts as a stable particle in this region.  
On the other hand, the physical state conditions, which 
include the condition for the Hamiltonian, 
$H |{\rm phys} \rangle =0$, seem to remove such an 
complex-energy state from the physical states, 
because it requires that the eigenvalue of the
Hamiltonian is real. To begin with, whether or not 
the usual particle picture can be 
preserved at this high-energy region is of concern. 
Here, there is a naive question: asymptotically free theories 
always have trivial asymptotic states in the high-energy region. 
This is true for four-dimensional Yang-Mills (4DYM) theories. 
However, it may not be true for 4DQG, because 4DQG is a 
four-derivative theory, so that we 
cannot neglect infra-red effects and the physical states may be not  
generated from free fields $\phi$ and $h^{\mu}_{~\nu}$ as usual 
Fock space of states.  

The spectrum of 4DQG is rather analogous to that of 2DQG 
composed to that of 4DYM.  
Remember that 2DQG is a free theory, but it has a rich 
structure of the physical states~\cite{bmi}, for example,
\bb
    \e^{2\phi}, \quad \pd X {\bar \pd} X, \quad 
    \Bigl( \pd^2 X \pm 2i (\pd X)^2 \Bigr) 
      \Bigl( {\bar \pd}^2 X \pm 2i ({\bar \pd} X)^2 \Bigr) 
        \e^{-\phi \mp iX}, \quad \cdots, 
\ee
where $X$ is a scalar field with central chage $c_X=1$. These infinite 
number of states are called discrete states. 
Note that there is an analogy between this two-dimensional field $X$ 
and the traceless mode, $h^{\mu}_{~\nu}$, in 4DQG, 
because $2n$-th order fields in $2n$ dimensions have a common infra-red 
behavior. As for the conformal mode,  
our theory has quite similar features to 2DQG in the sense 
that this mode is treated exactly, 
and the parts of the effective action related to this mode  
becomes scale-invariant.
Thus, the physical states of 4DQG may be given by composite fields: 
\bb
    \sq{g}, \quad \sq{g}R, \quad \sq{g}R^2, \quad \sq{g}F, 
    \quad \sq{g} F^2, \quad \cdots. 
\ee 
Hence, even at the $t_r \arr 0$ limit, 4DQG has a rich structure of 
the physical states. 

These asymptotic states are rather similar to glueball states in 4DYM, 
even though the theory is asymptotically free. 
Hence, to distinguish from a graviton, we here call them graviball states. 
On the other hand, the usual graviton state would be obtained  
as a Nambu-Goldstone mode when background-metric independence, 
namely $H |{\rm phys} \rangle =0$ condition, violates dynamically. 

The order parameter of this phase transition may be given by 
the vacuum expectation value (VEV) of $\sq{g}$. At the high-energy region, 
it is known that $\langle \sq{g} \rangle =0$ because the two-point 
function of $\sq{g}$ goes to zero with a power-law behavior of 
the distance measured by the background metric. 
This implies that the physical distance 
measured by the metric field vanishes. 
At low-energy, due to the dynamics of the traceless mode, the 
four-derivative terms are decoupled and background-metric independence 
is violated so that a non-zero value of $\langle \sq{g}\rangle$  
would be obtained.\footnote{
Note that the role of the coupling in 4DQG  
is different from that of 4DYM. 
In 4DYM, the order parameter is given by VEV of 
the temporal Wilson loop, $\langle W_t \rangle$, 
where $W_t=\e^{i\int A_0 dt}$. This order parameter vanishes  
in the confinment phase at the strong coupling, 
while in the weak coupling phase it has a non-zero value.  
} 
The concept of the distance would then arise.  

The physical states presented above are rather symbolic. 
When the gravitational effects are dominant compared to that of 
matter fields, we must include corrections, such as, for example  
$\sq{g} \arr \e^{\a \phi}$, where 
$\a=4+\gm_{\Lambda}$~\cite{kpz,ddk,am,h99}.  
The anomalous dimension of the cosmological constant is obtained 
by~\cite{h99}  
\bb
   \gm_{\Lambda} = \fr{\a^2}{4e_0} + t^2_r \biggl( 
        -\fr{e_1}{4}\fr{\a^2}{e_0^2} +\fr{7}{16}\fr{\a^2}{e_0} 
        +\fr{\a^2}{48} -\fr{1}{288}\fr{\a^3}{e_0} 
        +\fr{1}{512}\fr{\a^4}{e_0^2} \biggr) + o(t^4_r). 
\ee
A deviation from the classical value, $\a=4$, becomes large for a 
small value of $e_0$. 
 
Finally, we comment on recent interesting numerical 
results~\cite{bbkptt,efhty} 
in the  dynamical triangulation (DT) approach~\cite{migdal}. 
Let us expand $\a$ as $\sum_n \a_n t^{2n}_r$. 
We can fit the result of DT method with our theory approximated 
by a sufficiently small value of the coupling $t^2_r$~\cite{h99}. 
This is consistent with that the coupling is asymptotically free. 
The lowest term, $\a_0$, is now given by a function of $e_0$ only, 
so that from (\ref{evalue}), the matter-dependence is given by the 
function of the combination $N_X +62N_A$. 
Recently, this dependence is directly checked numerically 
by Horata, Egawa and Yukawa in DT approach~\cite{hey}.  
This is evidence that conformal anomalies play a crucial role in 4DQG.

\vspace{5mm}

\begin{flushleft}
{\bf Acknowledgements}
\end{flushleft}

I wish to thank H. Egawa, S. Horata and T. Yukawa for informing me of 
their recent interesting results~\cite{hey} before publication.  
This work is supported in part by the Grant-in-Aid for 
Scientific Research from the Ministry of 
Education, Science and Culture of Japan.

\end{document}